\documentstyle[epsfig]{aipproc}
\begin{document}
\title{What Have We Learned From GRB Afterglows?}
\author{J. I. Katz$^*$ and T. Piran$^\dagger$}
\address{$^*$Department of Physics and McDonnell Center for the Space
Sciences\\Washington University, St. Louis\\$^\dagger$Racah Institute of
Physics\\Hebrew University, Jerusalem, Israel}
\maketitle
\begin{abstract}
The discovery by BeppoSAX and coordinated ground-based observations of
persistent X-ray, visible and radio counterparts to GRB has successfully
concluded a search begin in 1973, and the observed redshifted absorption
lines have proved that GRB are at cosmological distances.  The problem of 
explaining the mechanisms of GRB and their persistent counterparts remains.
There are two classes of models: 1) GRB continue weakly for days at all 
frequencies; 2) GRB emission shifts to lower frequencies as relativistic 
debris sweeps up surrounding gas (in an ``external shock'') and slows.  1) 
predicts that the visible afterglow should be accompanied by continuing
gamma-ray emission, as hinted by the high energy emission of GRB940217 and 
the ``Gang of Four'' bursts of October 27--29, 1996.  It also suggests that
the persistent emission will fluctuate.  Behavior of this sort may be found
in ``internal shock'' models.  2) has been the subject of several 
theoretical studies which disagree in assumptions and details but which 
predict that at each frequency the flux should rise and then decline, with 
the maximum coming later at lower frequencies.  Some of this behavior has 
been observed, but data from GRB970508 show that its afterglow cannot be
simply extrapolated from its gamma-ray emission.  It is likely that both 
classes of processes occur in most GRB.  Comparisons between GRB show that 
they are not all scaled versions of the same event.  These results suggest 
that most gamma-ray emission is the result of ``internal shocks'' while most
afterglow is the result of ``external shocks'', and hint at the presence of
collimated outflows.  Self-absorption in the radio spectrum of GRB970508 
permitted the size of the radiating surface to be estimated, and in future 
GRB it may be possible to follow the expansion of the shell in detail and to
construct an energy budget.
\end{abstract}
\section*{Introduction}
     The visible, infrared and radio afterglows of GRB970228 and GRB970508
have taught us many things.  Some of them are obvious: The absorption
redshift $z = 0.835$ of GRB970508 established the cosmological distance
scale of GRB beyond any reasonable doubt, confirming the very strong case
made on statistical grounds\cite{jikm92} from BATSE data.  It is also
evident that afterglows are very faint.  This suggests that the simultaneous
visible counterparts to GRB, as yet unobserved, will also be faint, so that
experiments designed to detect them will have to be very sensitive.  The
loss of the original HETE, carrying an insensitive ultraviolet imager, may
therefore have been fortunate, for it will be replaced by HETE-2 which 
instead will carry a soft X-ray CCD which may yield important spectral 
information.

     Observations of afterglows can answer a number of harder questions too:
\begin{enumerate}
\item What is the relative importance of internal {\it vs.} external shocks?
\item Where does a GRB end and its afterglow begin?
\item Does gamma-ray activity last as long as the afterglow, and could it be
an inseparable part of the afterglow?
\end{enumerate}
These questions are central to the understanding and interpretation of
afterglows.  
\section*{Internal {\it vs.} External Shocks}
External shocks have been widely considered for GRB since they
were suggested by Rees and M\'esz\'aros\cite{jikrm92}.  They predict a
hard-to-soft evolution of the spectrum.  In many models the duration or
elapsed time $t$ is related to a characteristic emission frequency $\nu_c$ 
by a power law $t \propto \nu_c^{-\alpha}$; the exponent $\alpha$ is model 
dependent\cite{jikk94a,jikk94b,jikkp97} but is usually close to $1/2$.  This 
is in remarkably good agreement with X-ray and soft gamma-ray 
observations\cite{jikpi97} of the single-peaked GRB960720 in which $\alpha 
= 0.46$.  However, the fact that GRB with a wide range of durations (from
tenths to hundreds of seconds) have comparable $\nu_c$ argues against
external shock models, which generally predict that these two quantities
should vary roughly reciprocally; it is difficult for $\nu_c$ to be in the 
soft gamma-ray range if the duration is minutes.

There must be more to GRB than external shocks.  Fenimore, {\it et 
al.}\cite{jikfmn96} and Sari and Piran\cite{jiksp97} showed that an
external shock can produce only smoothly varying time-dependent emission, not
the spiky multi-peaked structure found in many (but not all) GRB.  Such
complex variation must reflect variation in the supply of energy; it cannot
be explained solely by interaction with a heterogeneous medium, however 
complex.  A variable outflow may radiate when different fluid elements 
interact with each other. This process is generally called an internal 
shock because it does not involve an external medium, although there need be
no shock in the hydrodynamic sense of a discontinuity in pressure, density 
and temperature between two fluids each in thermodynamic equilibrium.  
Kinematic constraints require that there be inelastic interaction between 
streams of matter emitted with widely differing Lorentz factors in order
that radiation be produced with reasonable efficiency.

Afterglows have, so far, been observed to have a smooth single-peaked time
dependence in visible light (their complex time dependence at radio
frequencies results from interstellar scintillation\cite{jikgo97}), and
therefore are naturally explained as a consequence of external shocks.  This
model predicted\cite{jikk94a,jikk94b} both the existence of afterglows and
the gradual rise to maximum which was observed in both
GRB970228\cite{jikgu97} and GRB970508\cite{jikpe98}.  However, there is
no evidence that internal shocks could not produce the required time
dependence, and they are therefore possible explanations for smooth
single-peaked GRB and for afterglows, as well as for the multi-peaked GRB
for which they are required.

Is it meaningful to distinguish an afterglow from the GRB itself?  This
question first arises for X-rays, whose photon energy range overlaps that
traditionally assigned to GRB.  It has no good answer, and
should be considered a matter of nomenclature rather than of physics.
Rather, we should be concerned with deciding which emission is produced by
internal shocks and which by external; the taxonomist may wish to label the
former the GRB and the latter the afterglow.

There are two possible limits\cite{jikkp97}.  In one all the radiation is
the product of an external shock.  This is excluded, at least for
multi-peaked GRB.  It may describe single-peaked GRB and their afterglows.
External shock models successfully predicted\cite{jikk94b} the delayed
maximum of visible afterglow brightness observed in
GRB970228\cite{jikgu97} and GRB970508\cite{jikpe98}, and are likely to
explain such afterglow emission even if they cannot explain gamma-ray
emission.  These models also successfully predicted\cite{jikk94a,jikkp97}
self-absorption in the radio afterglow, as was observed\cite{jikf97} from
GRB970508.

In the other limit all the emission of a GRB is the product of internal 
shocks, which may continue for days\cite{jikk97} beyond the nominal 
gamma-ray duration; the afterglow is a continuation, at lower intensity, of
the gamma-ray emitting phase.  This has been suggested for GRB970228, where
it agrees with the observed instantaneous spectrum\cite{jikkps97}, and the
predicted continuing X-ray emission (beyond the decaying afterglow) has been
observed from GRB970508\cite{jikpi98}.  This hypothesis is consistent with
the hours-long gamma-ray emission of GRB940217\cite{jikh94}, and may
explain the ``Gang of Four'' bursts of October 27--29, 1996\cite{jikcon97}
as a single GRB.
\section*{How to Interpret the Data}
     Early work on GRB and their afterglows attempted to construct analytic
and numerical models of the entire process.  Unfortunately, several
essential quantities are poorly known (and unlikely to be the same for all
GRB or uniform within a single GRB), such as the distribution of energy 
among the electrons, ions and magnetic field, the surrounding density and
magnetic field, the efficiency of radiation and the degree of
collimation.  As is so often the case, the real world has turned out to be
more complex than theorists could imagine, and a more phenomenological
approach is required.

Theories of radiation mechanisms generally predict instantaneous spectra, 
but observations of afterglows are rarely simultaneous across the spectrum; 
X-ray, visible and radio observers face different observational constraints.
It is possible to fit multiparameter models\cite{jikmr97,jikwrm97} to 
non-simultaneous data, but the phenomenologist would like a more direct 
comparison of the emission in different frequency bands.  For example, in 
each band of interest the peak spectral intensity can be measured, or at 
least estimated, if measurements are obtained close to and straddling the 
maximum.  The resulting function $F_{max}(\nu)$ is not an instantaneous 
spectrum, but makes it possible to compare emission mechanisms in different 
spectral bands.

\begin{figure}
\centerline{\epsfig{file=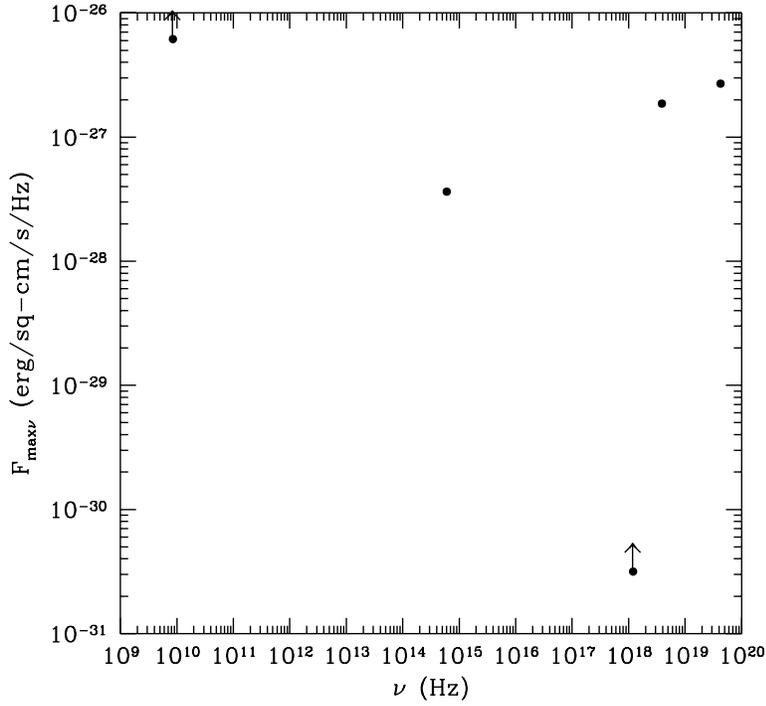,width=4.1in}}
\caption{$F_{max}(\nu)$ for GRB970508}
\label{jikfig1}
\end{figure}

In the case of GRB970508 $F_{max}(\nu)$ can be estimated at GHz, visible
(and near-IR), hard X-ray and soft gamma-ray energies (the maximum of the 
soft X-ray intensity was not observed), and is shown in Figure 
\ref{jikfig1}.  A single mechanism is unlikely to explain emission in all 
these bands, because $F_{max}(\nu)$ is not consistent with a single power 
law, as would be predicted by a single mechanism which does not have a 
characteristic frequency to define a spectral break.  The visible data form
a ``hinge''; on purely phenomenological grounds it is not possible to say 
whether this emission is produced by the same mechanism as the GHz emission 
or the hard X-rays and soft gamma-rays.  Neither of the limiting cases 
discussed in the previous section is satisfactory.  A plausible hypothesis 
suggests that GHz and visible emission result from external shocks and hard 
X-rays and soft gamma-rays (the classical GRB) from internal shocks.  This 
explains why an early model\cite{jikk94b} underestimated the time to maximum
and overestimated the brightness of the visible afterglow of GRB970508: it 
extrapolated from the gamma-ray emission of an assumed bright GRB, but this 
extrapolation was invalid (even if scaled to the lesser flux of GRB970508)
because the gamma-rays were produced by a different mechanism.

It is also possible to compare\cite{jikkp97} the properties of different GRB
and their afterglows by comparing the burst-to-burst ratios of 
$F_{max}(\nu)$ in different spectral bands.  The afterglow of GRB970508 was 
roughly 20 times brighter than that of GRB970228 in soft X-rays and visible 
light, when scaled to their soft gamma-ray brightnesses.  The afterglow of 
GRB970508 was at least four times brighter than that of GRB970111 at 1.43 
GHz, and at least 1000 times brighter than that of GRB970828 in visible 
light when similarly scaled (in these last two cases the ratios are lower 
limits because only upper limits exist to the fluxes of GRB970111 and 
GRB970828 in these bands).  These ratios quantify the conclusion that 
GRB970508 had a remarkably bright afterglow, compared to the the other GRB 
for which useful data exist, at all frequencies at which comparisons are 
possible.

This leads to the important (and unexpected) conclusion that afterglows are
``all different''.  They are not all scaled versions of the same event, and
any single simple model must fail.  It argues against models in which all 
GRB and afterglows are related by a single scaling parameter, such as the 
ambient density, energy or distance, or even some combination of these.  A
natural interpretation is that (at least in these GRB) the hard X-rays and 
gamma-rays are produced by internal shocks, and the lower frequency 
afterglow by an external shock.  If so, there need be no close correlation 
between the brightnesses of GRB and their afterglows, partly because the 
mechanisms are different, but also because internal shock properties depend
on the detailed temporal and spatial structure of the outflow and may well 
be very different in different GRB.
\section*{A Generic Afterglow Model}
A generic afterglow model is based on the assumption of an external shock,
which permits specific predictions to be made.  The interaction between a 
relativistic debris shell and the surrounding medium, or among various 
elements of an outflowing wind, has been the subject of many papers, but the
basic physics is not understood.  Even the essential collisionless shock is 
largely a matter of speculation, although recent calculations\cite{jikus97} 
have begun to attack the problem.  Still, a few features common to most 
afterglow models are independent of assumptions as to the mechanism of
entropy production.  The asymptotic ($\nu_s \ll \nu \ll \nu_c$) instantaneous
spectrum\cite{jikk94b} has the form $F_\nu \propto \nu^{1/3}$ between a 
self-absorption frequency $\nu_s$ and a characteristic synchrotron frequency
$\nu_c$.  Below $\nu_s$ the spectrum $F_\nu \propto \nu^2$, while above
$\nu_c$ the flux falls off with a slope reflecting the high energy ``tail''
to the particle distribution function; a power law is typically observed in
GRB, but its slope is unpredictable, and is observed to differ from burst to
burst and with time in a given burst.  The instantaneous spectra $F_\nu(t)$
are bounded from above by the function $F_{max}(\nu)$.  In simple analytic 
models $F_{max}(\nu) \propto \nu^{-\beta}$, with the exponent $\beta$ 
typically\cite{jikk94b,jikkp97} close to 0; any function of $\nu$ other than
a power law would define a characteristic break frequency and require a 
more complex model.

\begin{figure}
\centerline{\epsfig{file=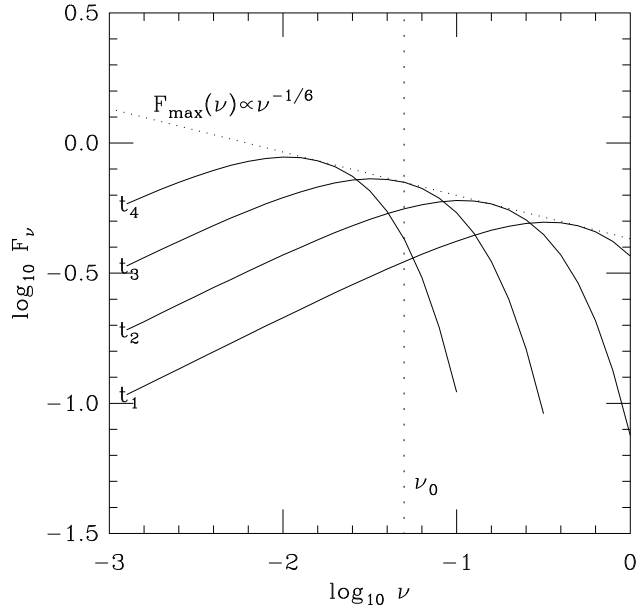,width=3.5in}}
\caption{Instantaneous GRB spectra for a model in which $F_{max}(\nu) 
\propto \nu^{-1/6}$; $t_1 < t_2 < t_3 < t_4$.}
\label{jikfig2}
\end{figure}

As time progresses both $\nu_s$ and $\nu_c$ decrease.  At a given frequency
$\nu_0$ the flux $F_{\nu_0}$ rises until $\nu_c$ declines to $\sim \nu_0$, 
after which $F_{\nu_0}$ decreases.  This predicted behavior has been 
observed, at least qualitatively, in the afterglows of 
GRB970228\cite{jikgu97} and GRB970508\cite{jikpe98}.  The predicted rate of
the initial rise $F_\nu(t) \propto t^\delta$, where the preceding 
expressions for $F_{max}(\nu)$ and $t(\nu_c)$ and the instantaneous $F_\nu$ 
lead to
$$\delta = {\beta + 1/3 \over \alpha}.$$
Typically, $\delta$ is estimated to be slightly less than unity; in one
model\cite{jikk94b} $\delta =$ 4/5 while in another\cite{jikkp97} $\delta =$
6/7.  Figure \ref{jikfig2} shows the evolution of the instantaneous spectrum.
$F_{\nu_0}$ rises until $t = t_3$ and then declines.
\section*{Testing the Generic Afterglow Model}
The predictions of the generic external shock afterglow model can be tested 
with data from the two observed afterglows.  The predicted rise to maximum
has been observed.  Some data\cite{jikd97} suggested agreement ($\delta 
\approx 0.7$) with the predicted rate of rise for GRB970508, but more 
complete data for GRB970228\cite{jikgu97} and GRB970508\cite{jikpe98} 
suggest a much steeper rise, with $\delta \approx 2$--3.

These large values of $\delta$ appear to disagree with the model.  
Sari\cite{jiksa97} pointed out that in the early stages of external shock
models, when the relativistic debris shell has not yet been significantly
slowed by the ambient medium, the luminosity should rise $\propto t^2$ at
all frequencies ($\delta = 2$), a simple consequence of the increasing
($\propto r^2$) area of the shell.  If this is the explanation then the
rapid rise should level out after a time
$$t \approx \left({3 E \over 4 \pi \rho c^2}\right)^{1/3} {1 \over 2 c
\gamma^{8/3}},$$
when deceleration becomes important (the earlier stages of the rising flux
could be hidden under steady emission by other processes, such as internal
shocks).  This leads to an estimate of the ambient density $\rho$:
$$\rho \approx 10^{-33} \left({E \over 10^{52}{\rm\,erg}}\right) \left(
{\gamma \over 10^2}\right)^8 \left({t \over 10^5 {\rm\,s}}\right)^{-3}\ 
{\rm g/cm}^3.$$
If the model is correct this implies an extraordinarily low ambient
density out to a radius $\sim 2 c \gamma^2 t \sim 20 (\gamma/10^2)^2
(t/10^5\,{\rm s})\,$pc, or a surprisingly low value of $\gamma \sim 10$.
Such a low density would be remarkable, even in the intergalactic medium
(although a pre-coalescence pulsar wind could create a bubble), and such a
low $\gamma$ is inconsistent with gamma-gamma pair production constraints
(although in a long duration internal shock model a low $\gamma$ wind could
follow a brief high $\gamma$ wind which produces the gamma-ray emission).

If the afterglow is produced by internal shocks it is probably not possible
to predict its time dependence.  However, the instantaneous spectrum is
still predictable.  There is some spectral evidence that the first several
hours of afterglow in GRB970228 were the product of internal
shocks\cite{jikkps97}.  Internal shocks are a possible explanation of
disagreements between the observed time dependence and predictions of the 
generic external shock model.

This early (3--8 hours after the GRB) period of roughly constant visible
intensity in GRB970508 poses another problem.  In either internal or 
external shock models, if the electron synchrotron cooling time is short 
compared to the duration of observation then the observed spectrum is that 
integrated over the electrons' cooling history\cite{jikco97}: $F_\nu 
\propto \nu^{-1/2}$.  Comparison of the visible and soft
X-ray\cite{jikpe98} fluxes during this period shows a deficiency of X-rays
compared even to this spectrum, suggesting that $\nu_c$ lies within or below
the X-ray band.  This is consistent with emission by an internal shock with 
a low value of $\gamma$, as suggested by both the near constant visible 
intensity and the long-delayed onset of the rapid rise.

These conclusions are based on limited data from two afterglows.  The outline
of the external shock afterglow model is supported by the data, but the 
detailed interpretation, especially of the rise of the visible intensity to 
maximum, must await more data from more afterglows.
\section*{The Instantaneous Spectrum}
The predicted\cite{jikk94b} instantaneous asymptotic spectrum $F_\nu \propto
\nu^{1/3}$ for $\nu \ll \nu_c$ should be applicable to both internal and 
external shocks.  It is based on several plausible assumptions: incoherent 
synchrotron radiation, no cooling (radiative or adiabatic) and a phase space
argument for the electron distribution function produced by a relativistic 
shock which heats the entire electron distribution function, rather than
just a ``tail'' of suprathermal particles.  It should apply to
the synchro-Compton spectrum too, but with a different coefficient.

Like all theoretical predictions, it is only speculation until empirically
tested.  So far, the data are inconclusive and not completely consistent.
Some X-ray and soft gamma-ray observations of GRB support the
prediction\cite{jikco97,jiks97}, but others disagree\cite{jikpr98,jikclp98}.
The inconsistency may result from the difficulty of extracting quantitative 
spectral information from NaI scintillator data, which have low intrinsic 
spectral resolution, especially at photon energies $<100$ KeV where the 
asymptotic spectrum is expected.  This difficulty may also account for the 
long-standing controversy over the reality of line features in GRB 
radiation, and may only be resolved when data from detectors of 
intrinsically higher resolution become available.

Visible data\cite{jikgr98} lead to an exponent $0.25 \pm 0.25$ in the
pre-maximum phase of the afterglow of GRB970508, in agreement with the
predicted 1/3 (agreement is not expected at and after maximum, because then
the frequency of observation exceeds $\nu_c$).  Radio data\cite{jikf97}
from the same afterglow lead to an exponent $\approx 0.2$, also in agreement
with prediction.  The results are encouraging, but not yet conclusive.

If a GRB were to be detected in visible light during its initial brief phase
of gamma-ray emission the spectral exponent could be determined quite
accurately by comparing fluxes at frequencies separated by a factor of
more than $10^4$.  Such simultaneous detection remains the holy grail of GRB 
visible counterpart research.
\section*{Self-Absorption}
Self-absorption of GRB afterglows at GHz frequencies was 
predicted\cite{jikk94a,jikkp97} and confirmed\cite{jikf97}.  This was no
great surprise; the physics is elementary, though a little different (the
spectral exponent below $\nu_s$ is predicted to be 2 rather than 2.5) than
in the usual case of synchrotron self-absorption by a power law distribution
of electron energies.  Self-absorption is important because it may lead to a
measurement of the emission radius $r$ as a function of time with few
uncertain assumptions.  The flux for $\nu \ll \nu_s$ is
$$F_\nu = 2 \pi \nu^2 m_p \zeta (1 + z) {r^2 \over D^2},$$
where $D$ is the distance, $z$ the cosmological redshift, $m_p$ the proton
mass and $\zeta$ an equipartition factor defined as $k_B T_e / \gamma m_p 
c^2$; $\zeta = 1/9$ in the case of complete electron-ion-magnetic 
equipartition.  The shock Lorentz factor (assumed $\gg 1$) drops out; this 
is fortunate, for it is poorly known and likely to remain so.

If $r(t)$ were inferred from measurements of the self-absorbed flux then it
would be possible to reconstruct the expansion history and slowing down of
the relativistic debris in an external shock model.  This would permit
determination of the ambient density and the efficiency of radiation as the
initial kinetic energy is radiated or shared with swept-up matter.
A preliminary attempt\cite{jikkp97} to construct such an energy budget
(based on one inferred value of $r$) for the afterglow of GRB970508 led to
the conclusion that the total kinetic energy after seven days was only $\sim
10^{49}n_1$ erg, where $n_1$ is the ambient particle density.  This is much
less than the $\sim 3 \times 10^{51}$ erg inferred from the gamma-ray
radiation, assuming isotropic emission.  

There are three possible implications of this result: $n_1 \gg 1$ cm$^{-3}$,
as might be found in a dense star-forming region; strong beaming of the
gamma-rays (and therefore of the initial relativistic outflow); or a 
radiation efficiency $>99$\% in the first seven days of the event.  Any or 
all of these are possible, and any or all would be important.
\section*{A Theorist's Wish List}
We have compared the observed afterglows to theoretical predictions based on
simple models, and have generally found qualitative agreement.  In order to
test the theory, particularly that of relativistic shocks, in more detail,
more difficult measurements will have to be performed.  Here is a list of
ambitious goals a theorist might set for his observational colleagues:
\begin{enumerate}
\item Accurate X-ray spectroscopy at moderate resolution ($\sim 30$) to test
the predicted instantaneous $F_\nu \propto \nu^{1/3}$ ($F_\nu \propto 
\nu^{-1/2}$ if synchrotron cooling is important).  This may be performed by 
ZnCdTe or Ge detectors, or by X-ray CCD.
\item Simultaneous spectral measurements across several decades of
frequency, with $\nu < \nu_c$ throughout (hence the intensity must not have
reached its maximum anywhere in this range).  This could be achieved by 
observation of the visible counterpart to a GRB during its strong gamma-ray
activity, or by simultaneous measurements of visible and radio afterglow 
before the visible maximum (within the first day for GRB970228 and GRB970508).
\item Measurement of the intensity as a function of time in the
self-absorbed regime of radio afterglow.
\end{enumerate}
Each of these measurements is likely to be difficult.  Locating and
measuring a visible counterpart to a GRB within tens of seconds is much
harder than doing it with several hours of imaging X-ray observations.  The
only afterglow observed at radio frequencies was not strong enough to be
detected until long after the visible maximum, and self-absorption reduces
the strength of the radio emission even more.  The first goal is probably
the most feasible, and would test relativistic shock theory; the last is
probably the one which would lead to the greatest understanding of how GRB and
their afterglows really work.
\section*{Acknowledgements}
We thank R. Sari for discussions.  JIK thanks NSF 94-16904 for support, 
Washington University for the grant of sabbatical leave and the Hebrew 
University for hospitality and a Forchheimer Fellowship.

\end{document}